# Discrete derivative estimation in LISA Pathfinder data reduction

**Luigi Ferraioli, Mauro Hueller and Stefano Vitale**

Dipartimento di Fisica and INFN, Università di Trento, Trento, Italy

E-mail: luigi@science.unitn.it

**Abstract.** Data analysis for the LISA Technology package (LTP) experiment to be flown aboard the LISA Pathfinder mission requires the solution of the system dynamics for the calculation of the force acting on the test masses (TMs) starting from interferometer position data. The need for a solution to this problem has prompted us to implement a discrete time domain derivative estimator suited for the LTP experiment requirements. We first report on the mathematical procedures for the definition of two methods; the first based on a parabolic fit approximation and the second based on a Taylor series expansion. These two methods are then generalized and incorporated in a more general class of five point discrete derivative estimators. The same procedure employed for the second derivative can be applied to the estimation of the first derivative and of a data smoother allowing defining a class of simple five points estimators for both. The performances of three particular realization of the five point second derivative estimator are analyzed with simulated noisy data. This analysis pointed out that those estimators introducing large amount of high frequency noise can determine systematic errors in the estimation of low frequencies noise levels.

## 1. LTP Dynamics

Among the main goals of the LTP experiment, to be flown aboard the LISA PATHFINDER mission, is the estimation of the residual force noise acting on the test masses, perturbing them from their geodesic motion. The measurement is performed in a differential configuration, by looking at the relative acceleration of two test masses hosted on the same spacecraft (SC) [1 – 4]. In the framework of the first LTP Mock Data Challenge (MDC) [1], the differential acceleration of the test masses was reconstructed by the output of the interferometric readout according to the equation:

$$\begin{aligned} a_{n1}(s) &= \left(s^2 + \omega_{p1}^2\right)o_1(s) + H_{df}(s)o_1(s) \\ a_{n\Delta}(s) &= \left[\omega_{p2}^2 - \omega_{p1}^2 - \delta\left(s^2 + \omega_{p2}^2\right)\right]o_1(s) + \left(s^2 + \omega_{p2}^2\right)o_\Delta(s) + h_{lfs}(s)o_\Delta(s) \end{aligned} \qquad (1)$$

where:

- $o_1$ and $o_\Delta$ are the output of the interferometer channels reading the position of test mass 1(TM1) and the differential position between TMs respectively.
- $H_{df}(s)$ and $h_{lfs}(s)$ are the gains of the drag-free loop and low-frequency-suspension loop respectively [3, 4].
- $\omega_{p1}^2$ and $\omega_{p2}^2$ are the residual coupling of test mass 1 and test mass 2 (TM2) to the SC.
- $\delta$ is the cross-talk coefficient between the two interferometer (IFO) channels.

In writing Equation (1) we neglected small terms associated with the gravitational gradient and the mass ratio between TMs and S/C.

Acceleration of the two test masses is calculated by differentiating the interferometer outputs $o_1$ and $o_4$ taking into account the effects from residual coupling of the TMs to the SC and the presence of digital control loops applied to the spacecraft (following test mass 1 by means of low noise microthrusters) and to test mass 2 (forced electrostatically to follow TM1).

Discrete representation of the controllers was obtained by impulse-invariant discretization of their Laplace domain representations $H_{df}(s)$ and $h_{lfs}(s)$. Available IFO signals are discrete therefore the evaluation of the derivatives involves a discretized differentiation of the signals, which is obtained by suitably designed filters that mimic the frequency response of the derivative operators.

The subsequent step in estimating the residual noise is evaluating the Power Spectral Density (PSD) of the relative acceleration of the TMs, which is performed employing the modified Welch periodogram, associated with windowing and detrending of the time series to suppress long-term drifts. We discuss here the implications on these low frequency estimates as long as different methods are employed for signal differentiation. The most sensitive frequency range for LISA will be $1x10^{-3} - 3x10^{-2}$ Hz. Therefore the need to extract as much information as possible from available data demand for an accurate estimation of the components of the acceleration PSD in the frequency range mentioned above.

## 2. Derivative estimation

Given a discretized series of data points one of the possible ways to numerically estimate the derivative is to making use of a five point equation. The advantage of such estimators is to ensure high accuracy with reduced calculation efforts [5]. The five point method for the derivative approximation, at a given time $t = kT$ ($k$ being an integer and $T$ being the sampling time), is calculated by means of finite differences between the element at $t$ with its four neighbors. In other words, considering the discrete series of data:

$$\{y[1], y[2], \ldots, y[k], \ldots\} \tag{2}$$

First and second derivative at a certain time can be approximated by a five point difference equation:

$$\begin{aligned}
\frac{dy[k]}{dt} &\approx \frac{a'y[k-2]+b'y[k-1]+c'y[k]+d'y[k+1]+g'y[k+2]}{T} \\
\frac{d^2 y[k]}{dt^2} &\approx \frac{a''y[k-2]+b''y[k-1]+c''y[k]+d''y[k+1]+g''y[k+2]}{T^2}
\end{aligned} \tag{3}$$

where T is the sampling time. The problem of the identification of the derivative estimator is reduced to the definition of the five coefficients $[a', b', c', d', g']$ and $[a'', b'', c'', d'', g'']$ for the first and second derivative respectively.

In the framework of the first LTP MDC [1], we have used two different identification methods. The first is based on a parabolic fit approximation, and the second is based on a Taylor series expansion. In the following, it will be demonstrated that these two methods can be considered as two particular cases of a general method for the definition of a five point stencil for the numerical derivative estimation.

### 2.1. Parabolic fit approximation

This method estimates the coefficients of the difference equation by means of a least square fit with a second order equation on a generic data series:

$$y[(k+m)T] = \alpha_0[k] + \alpha_1[k]mT + \alpha_2[k](mT)^2 \tag{4}$$

Given the equation (4), it is easy to see the connection between fit coefficients and the derivative estimator:

$$y[t]_{t=kT} \approx \alpha_0, \quad \left(\frac{dy[t]}{dt}\right)_{t=kT} \approx \alpha_1, \quad \left(\frac{d^2y[t]}{dt^2}\right)_{t=kT} \approx 2\alpha_2 \qquad (5)$$

The least square procedure is equivalent to solve the system of equations:

$$\begin{pmatrix} y[(k-2)T] \\ y[(k-1)T] \\ y[(k)T] \\ y[(k+1)T] \\ y[(k+2)T] \end{pmatrix} = \begin{pmatrix} 1 & -2T & (-2T)^2 \\ 1 & -T & (-T)^2 \\ 1 & 0 & 0 \\ 1 & T & T^2 \\ 1 & 2T & (2T)^2 \end{pmatrix} \begin{pmatrix} \alpha_0 \\ \alpha_1 \\ \alpha_2 \end{pmatrix} \rightarrow \vec{Y} = \vec{X} \cdot \vec{B} \qquad (6)$$

The vector of coefficients can be found from the solution of $\vec{B} = (\vec{X}^T \cdot \vec{X})^{-1} \cdot \vec{X}^T \cdot \vec{Y} = \vec{K} \cdot \vec{Y}$, where $K$ matrix is:

$$\vec{K} = \begin{pmatrix} -\frac{3}{35} & \frac{12}{35} & \frac{17}{35} & \frac{12}{35} & -\frac{3}{35} \\ -\frac{1}{5T} & -\frac{1}{10T} & 0 & \frac{1}{10T} & \frac{1}{5T} \\ \frac{1}{7T^2} & -\frac{1}{14T^2} & -\frac{1}{7T^2} & -\frac{1}{14T^2} & \frac{1}{7T^2} \end{pmatrix} \qquad (7)$$

It is worth to note that $K$ is independent from the particular series of data. Comparing (7) with equation (5), we see that the first row of $K$ matrix is the vector of coefficients for the five point estimator of y[k] acting like a smoother for the data series, the second row of $K$ matrix is the vector of coefficients for the five point estimator of the first derivative of y[k] and the third row of $K$ matrix is the vector of coefficients for the five point estimator of one half the second derivative of y[k]. This method was successfully employed to reconstruct the external force acting on a harmonic oscillator as described in [6].

*2.2. Taylor series approximation*
The Taylor series expansion method for the determination of the coefficients of the five point derivative estimator is based on the 4$^{th}$ order expansion in Taylor series of y[(k+n)T], as it is considered as a generic function of (k+n)T:

$$y[(k+n)T] \approx y[kT] + \frac{dy}{dt}\bigg|_{n=0} nT + \frac{1}{2}\frac{d^2y}{dt^2}\bigg|_{n=0}(nT)^2 + \frac{1}{6}\frac{d^3y}{dt^3}\bigg|_{n=0}(nT)^3 + \frac{1}{24}\frac{d^4y}{dt^4}\bigg|_{n=0}(nT)^4 \qquad (8)$$

By means of equation (8) we can construct a system of four equations with $n = \pm 2, \pm 1$. Solving out for the first, second, third and fourth derivative of $y$, the first to fourth derivative estimators can be found [7]. Corresponding coefficients are:

$$\begin{aligned}
\frac{d}{dt} &\to \frac{1}{T}\left[\frac{1}{12},\ -\frac{8}{12},\ 0,\ \frac{8}{12},\ -\frac{1}{12}\right] \\
\frac{d^2}{dt^2} &\to \frac{1}{T^2}\left[-\frac{1}{12},\ \frac{16}{12},\ -\frac{30}{12},\ \frac{16}{12},\ -\frac{1}{12}\right] \\
\frac{d^3}{dt^3} &\to \frac{1}{T^3}\left[-\frac{1}{2},\ 1,\ 0,\ -1,\ \frac{1}{2}\right] \\
\frac{d^4}{dt^4} &\to \frac{1}{T^4}[1,\ -4,\ 6,\ -4,\ 1]
\end{aligned} \quad (9)$$

*2.3. Five point generalized method*

We now present a procedure for the estimation of a general expression for the derivative estimator coefficients that incorporates the two precedent methods as particular cases.

*2.3.1. Second order derivative.* Starting from equation (3), we calculate the Z transform of the general five point estimator for the second derivative as:

$$\frac{d^2}{dt^2} \to \frac{1}{T^2}\left[a''z^{-2} + b''z^{-1} + c'' + d''z + g''z^2\right] \quad (10)$$

Introducing the angular frequency $\Omega = 2\pi f/f_s = \omega T$, the frequency response of the estimator can be found evaluating equation (10) on the unit circle of the complex plane ($z \to e^{i\Omega}$):

$$\frac{d^2}{dt^2} \to \frac{1}{T^2}\left[(a''+g'')\cos(2\Omega) + i(g''-a'')\sin(2\Omega) + (b''+d'')\cos(\Omega) + i(d''-b'')\sin(\Omega) + c''\right] \quad (11)$$

We also know that the Laplace equivalent for the differential operator $d^2/dt^2$ in terms of angular frequency is:

$$\frac{d^2}{dt^2} \to s^2 \to -\omega^2 \to -\frac{\Omega^2}{T^2} \quad (12)$$

Comparing equations (11) and (12) we can make some considerations on the coefficients of the derivative estimator:

i) $\Omega^2$ in equation (12) is a real variable so we can expect the expression in square brackets of equation (11) being real too. The reality condition entails the vanishing of the imaginary part in expression (11). This happens if:

$$g'' = a'',\ d'' = b''\ \text{and}\ c'' \in R \quad (13)$$

ii) The continuous operator in equation (12) goes to zero when $\Omega \to 0$. We can expect the same behavior also for the numerical estimator:

$$\lim_{\Omega \to 0}\left[2a''\cos(2\Omega) + 2b''\cos\Omega + c''\right] = 0 \Rightarrow c'' = -2(a'' + b'') \quad (14)$$

iii) We can also impose the frequency response of the numerical estimator being a good approximation of the continuous operator at least for small values of $\Omega$. This is important because the focus of the LTP experiment is indeed the study of the disturbance forces acting on the TMs in the low frequency ($1x10^{-3}$ – $3x10^{-2}$ Hz) range which will be the most sensitive for LISA. Expanding in Taylor series up to the second order $\cos(2\Omega)$ and $\cos(\Omega)$, we can write:

$$2a''(1-2\Omega^2) + 2b''\left(1-\frac{\Omega^2}{2}\right) + c'' = -\Omega^2 \Rightarrow b'' = 1 - 4a'' \qquad (15)$$

Therefore all the coefficients in equation (3) can be written as a function of one of them (in this case we have chosen $a''$ as the independent coefficient):

$$\begin{aligned} b'' &= 1 - 4a'' \\ c'' &= -2(a'' + b'') = -2(a'' + 1 - 4a'') = -2(1 - 3a'') \\ d'' &= b'' = 1 - 4a'' \\ g'' &= a'' \end{aligned} \qquad (16)$$

The parabolic fit approximation and the Taylor series expansion methods can be found as particular cases of this general method if one sets $a'' = 2/7$ and $a'' = -1/12$ respectively. The frequency response of these two methods is reported in figure 1. Comparing them to the continuous estimator $-\Omega^2$ it can be seen that the Taylor series estimator is accurate (difference between theory and estimator lower than 10 %) up to $\Omega = 1.9$ (rad) whereas the parabolic fit estimator is accurate up to $\Omega = 0.6$ (rad). Parabolic fit estimator has also a zero at 2.42 (rad) this occurring as a notch feature in the spectrum of filtered data. Due to its higher accuracy, the Taylor series estimator has the undesirable characteristic of increasing the high frequency content in the filtered data so introducing a systematic error in the estimation of low frequency properties (details will be discussed in paragraph 4). From the other side the notch feature introduced by the parabolic fit estimator can increase the difficulty of modeling the spectral data. For these reasons we proposed also a second derivative estimator characterized by a zero at the Nyquist frequency (we called it PI). Its coefficients can be obtained setting $a'' = 1/4$ in equation (16). The frequency response of this estimator is reported in figure 1, it has the advantage of keeping down high frequency noise and it is free from notch features in the whole usable frequencies range.

*2.3.2. First order derivative.* In analogy to the second derivative case, we can start from equation (3) and write the Z transform of the five point first derivative estimator:

$$\frac{d}{dt} \rightarrow \frac{1}{T}\left[a'z^{-2} + b'z^{-1} + c' + d'z + g'z^2\right] \qquad (17)$$

First derivative operator in Laplace notation is $s$ whose frequency response is a pure imaginary quantity $i\omega$. Comparing the frequency response of equation (17) with that of the Laplace operator we can extract a set of conditions for the coefficients of the derivative estimator. The first information comes from the imaginary property of the derivative operator which allow us to set $g' = -a'$ and $b' = -d'$. Then we force the estimator going to zero for $\Omega \rightarrow 0$, this add the condition $c' = 0$. We are interested to a good estimator for the low frequencies region so we impose its response being a good approximation to the continuous operator at least to the second order in $\Omega$. From this last condition we obtain $b' = -1/2 - 2a'$. These are a set of conditions allowing the expression of the coefficients in equation (3) in terms of the first one $a'$. In particular the parabolic fit coefficients and the Taylor series expansion coefficients can be recovered by setting $a' = -1/5$ and $a' = 1/12$ respectively.

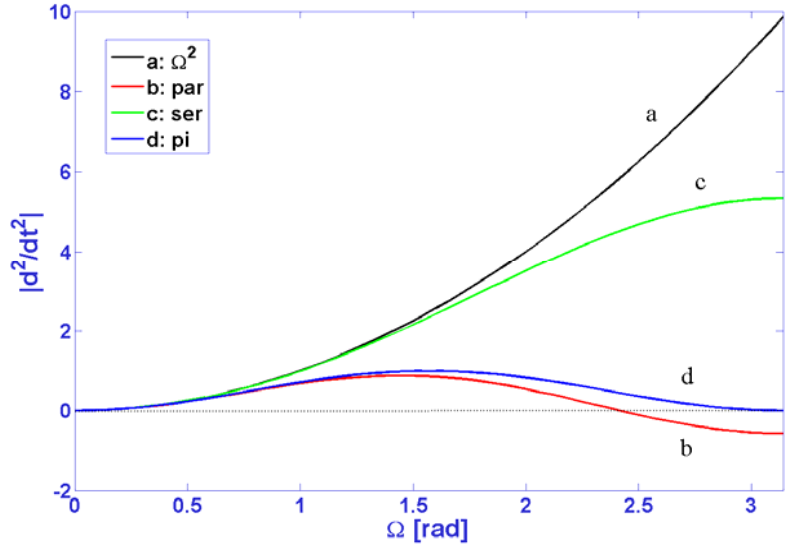

**Figure 1:** Angular frequency response of three different second derivative estimators. Red curve (b) refers to the parabolic fit estimator, green curve (c) refers to the Taylor series expansion estimator and blue curve (d) refers to the estimator with zero at the Nyquist frequency (PI). Black line (a) refers to the frequency response of the Laplace equivalent of the second derivative operator.

## 3. Five point data smoother

As it is shown in paragraph 2.1, a smoother can be obtained in association with the derivative estimator using the five point parabolic fit approximation. Regarding the parabolic fit estimator, the smoother and the differentiator have to be used together in the calculation of the dynamical equation because they are both obtained by a $\chi^2$ minimization procedure. On the other hand a general expression for the five point smoother can be found in analogy to what we have done for the first and second derivative. We are searching a set of five coefficients for the estimation of data points in a series:

$$y[k] \approx ay[k-2] + by[k-1] + cy[k] + dy[k+1] + gy[k+2] \tag{18}$$

Passing to the Z transform of (18) one finds that the spectral response of the estimator needs to be real so resulting in the condition $g = a$ and $d = b$. Then we can consider the estimator should be as much as equal to 1 on the whole frequency range. Forcing this equality at $\Omega = 0$ we obtain $2a + 2b + c = 1$. At this point, we can define the low pass nature of the estimator by forcing it to unity for small values of $\Omega$ and not only at zero. The final condition allows to defining the relation between the different coefficients as:

$$\begin{aligned} g &= a \\ d &= b \\ b &= -4a \\ c &= 1 + 6a \end{aligned} \tag{19}$$

## 4. Discussion with simulated data

In order to give more insight to the properties of the derivative estimators, we generated a set of noisy data corresponding to the displacement fluctuation of the TMs and characterized by a $1/f^2$ behavior at low frequency and a white plateau at high frequency according to the equation:

$$PSD = \frac{A}{f^2} + B \quad [m^2/Hz]. \tag{20}$$

Where $A = 10^{-9}$ m$^2$Hz and $B = 10^{-4}$ m$^2$/Hz. In the model, $1/f^2$ noise starts to be dominant for frequency values below $10^{-3}$ Hz (ratio between $A/f^2$ and $B$ larger than 10). We generated a time series, $10^5$ seconds long with a sampling frequency of 10 Hz and we also added a linear trend in time in order to simulate real measurement conditions. Data generation and spectra calculation were performed with the data analysis tools provided by the LTPDA Toolbox [2] running under MATLAB [8].

Power spectral density was calculated with the standard averaged Welch periodogram method using a 4-term Blackman-Harris window [9]; an overlap of 66% was used for averaging on 27 data segments. The choice of such a window is justified by our requirements in terms of spectral leakage performances. 4-term Blackman-Harris window, having the highest side-lobe level of -92 dB (relative to the main lobe level) [9], is indeed one of the best-performing available window in terms of spectral leakage suppression.

Once we had a time series of data we used the coefficients reported in equation (16) to calculate the second derivative of it. In particular we made use of three different values for $a''$ corresponding respectively to the parabolic fit method (parfit), the Taylor series method (Taylor) and the PI method (with the notch at the Nyquist frequency). Final results are showed in figure 2 where the spectra obtained with the three estimators are reported and compared with the theoretical expectation. Calculation of spectra was performed with and without the application of a linear fit detrend for each windowed data segment. The notch features of parfit and of PI methods can be easily identified; the first being located at 3.85 Hz and the second at 5 Hz corresponding to the Nyquist frequency.

In order to quantitatively discuss the methods accuracy we reported in figure 3 the comparison of the power spectral densities for the three methods (no detrend) with theoretical expectation. 95 % confidence bounds were calculated according with the procedure reported in [10, 11] and and reported in the figure as dashed light colored lines. We also see that in the frequency range between $2x10^{-3}$ Hz and $5x10^{-1}$ Hz the three methods are practically equivalent and in good agreement with the theoretical predictions within the confidence levels. Taylor series method can be considered accurate up to 4 Hz, above this frequency the deviation from the expected values is so large to make the estimated values inconsistent with the expected ones. Parfit and PI methods instead start to be inconsistent with theoretical expectation already for frequencies higher than 1 Hz.

As previously stated we are particularly interested to the low frequency region in the LTP experiment ($10^{-3}$ – $3x10^{-2}$ Hz). In account of this, the three methods could be considered equivalent because their responses in the frequency range of interest are consistent with theoretical expectations. A careful analysis of the frequency range below $10^{-3}$ Hz of the power spectra in figures 2 and 3 instead, clearly highlights a remarkable departure of the noise level from the theoretical expectation in the case of the Taylor estimator (figure 2c and 3). As a matter of fact, the spectrum relative to the Taylor method become inconsistent with the theoretical expectation below $10^{-3}$ Hz while Parfit and PI got to be inconsistent below $4x10^{-4}$ Hz. We also notice that the detrend operation introduces a slight systematic error in the values of the very low frequency noise in any case.

The reason for the loss of accuracy of the Taylor estimator is connected with its property of introducing large quantities of high frequency noise in comparison with parfit and PI estimators. As a matter of fact, high frequency noise is transferred during the spectrum calculation to the lowest frequency bins so determining a systematic error in the low frequency noise estimation. This assertion is supported by the spectral curves reported in figure 4 where the spectra of low passed derivative time series are reported. The filter used was a simple low pass infinite impulse response (IIR) filter with cutoff frequency at 0.5 Hz. The estimation of the low frequency noise level is highly improved for the Taylor method but parfit and PI methods remain still more accurate than Taylor.

It is worth to note that the ripple features, clearly visible in the spectra of figures 2 – 4, can be attributed to random fluctuation of the spectral estimation. They can be in fact effectively reduced by averaging the spectra for different and independent realizations of the same physical process.

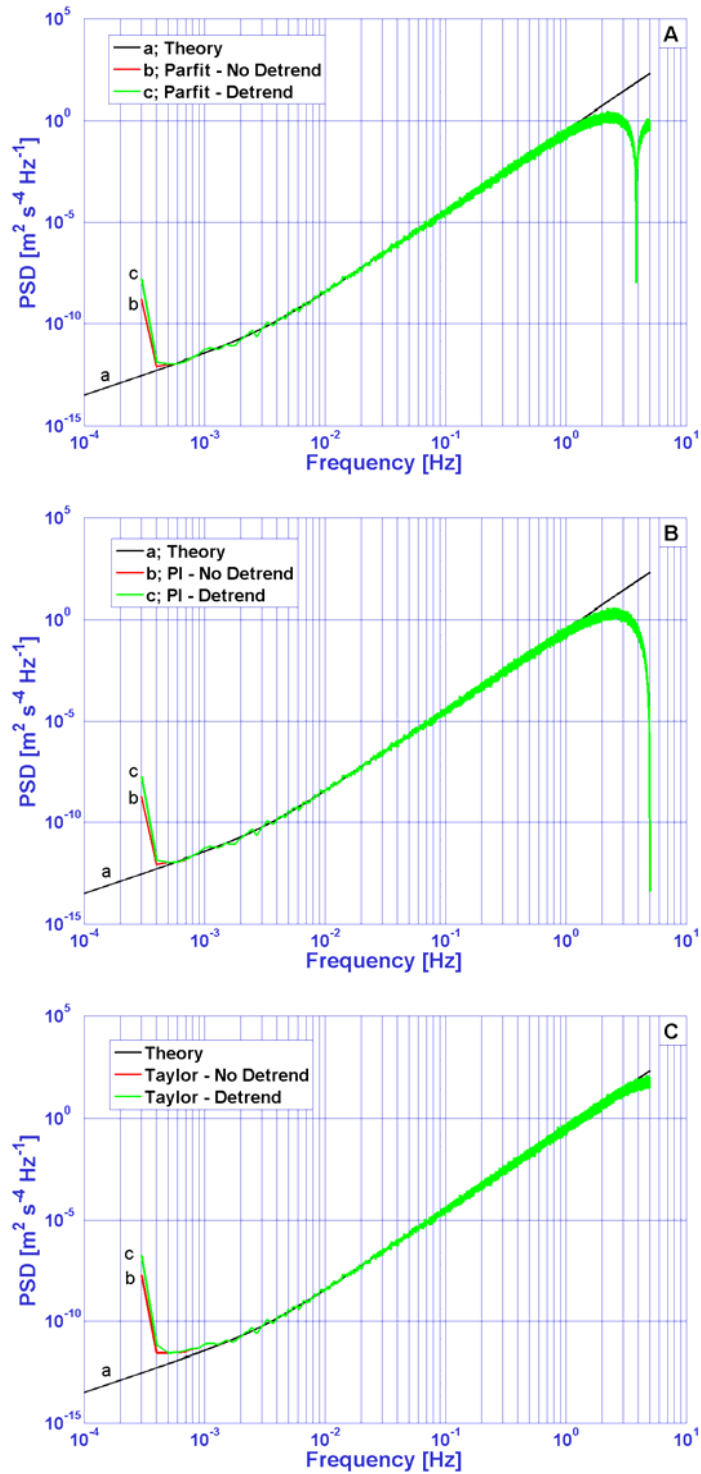

**Figure 2:** Power spectral density of simulated data after differentiation with the parfit (A) method, PI (B) method and Taylor (C) method. Black (a) line refers to the theoretical expectation. Red (b) line refers to the spectra calculated without performing the detrend of data. Green (c) line refers to the spectra calculated performing a linear fit detrend on each of the 27 windowed segment of data used for the Welch averaging procedure.

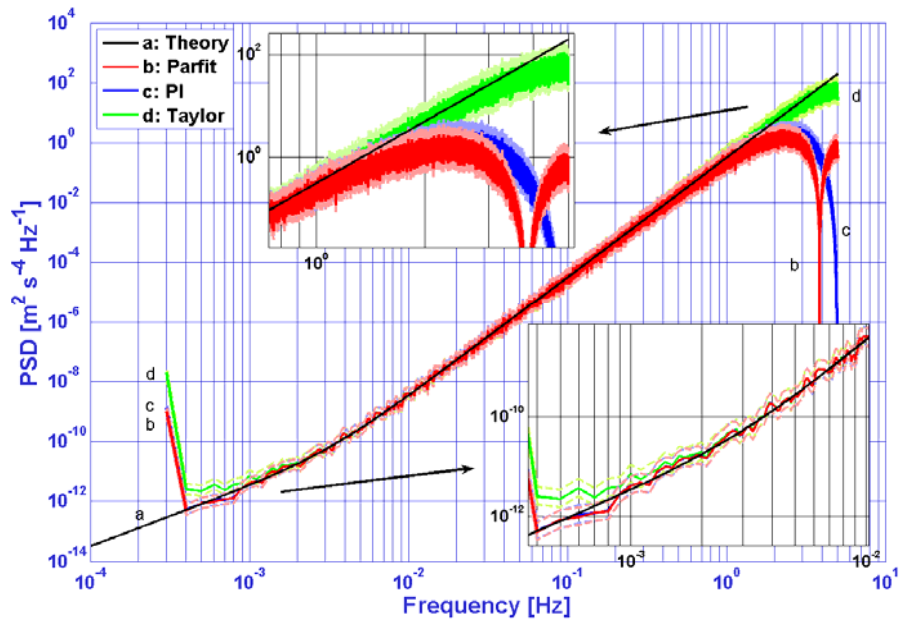

**Figure 3:** Power spectral density of simulated data differentiated with the three derivative methods and compared with the theoretical expectation. 95 % confidence bounds were calculated in accordance with the procedure described in [10, 11] and reported in the figure as dashed light colored lines. The inserts in figure highlight the low and high frequency regions were the difference between the three derivative methods is sensibly evident.

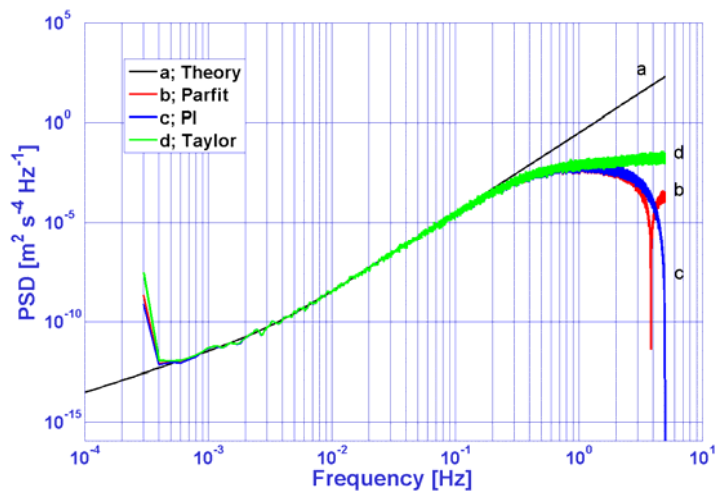

**Figure 4:** Power spectral density of low passed data after the differentiation process. (a) Black line corresponds to the theoretical expectation (without low pass), (b) red line corresponds to the parabolic fit estimator, (c) blue line corresponds to PI estimator and (d) green line corresponds to Taylor series estimator.

## 5. Conclusions

We reported on the estimation of the time domain numerical derivative with a five point formula for the dynamical calculations in the LISA Pathfinder data analysis. Starting from two, in appearance, different methods (parabolic fit and Taylor series expansion) we were able to define a general derivative estimator encapsulating them as special cases. This procedure was applied firstly to the second derivative and then extended to the first derivative and to the data smoother. The performances of three different second derivative estimators were analyzed with the help of a set of simulated data. The analysis of the spectra has evidenced that low frequency noise data could be corrupted by the transfer of high frequency noise during the spectra calculation procedures. The effect is particularly evident when noisy data are differentiated with those estimators that introducing a substantial amount of high frequency noise (such as the Taylor estimator). This effect can be substantially reduced low passing the data after the differentiation procedure even though estimators like parfit or PI, having an intrinsically low passing behavior, always ensure a more accurate low frequency noise level estimation as can be seen in figure 4 at $4x10^{-4}$ Hz.